\documentclass[12pt,aps,preprint]{revtex4}

\usepackage{epsfig}
\usepackage{amssymb}
\usepackage{amsmath}
\usepackage{amsfonts}
\usepackage{graphics}

\newcommand{\mtrx}[2]{\left(\begin{array}{#1} #2 \end{array}\right)}
\newcommand{\I}[0]{{\rm i}}
\newcommand{\eps}{\varepsilon}
\newcommand{\Refs}{Refs.}
\newcommand{\Ref}{Ref.}
\newcommand{\Sec}{Sec.}
\newcommand{\tab}{Table}
\newcommand{\eq}{Eq.}
\newcommand{\eqs}{Eqs.}
\newcommand{\fig}{Fig.}
\newcommand{\figs}{Figs.}

\newcommand{\ie}{\emph{i.e.}}
\newcommand{\eg}{\emph{e.g.}}

\newcommand{\CP}{\emph{CP}}

\DeclareMathOperator{\diag}{diag}
\DeclareMathOperator{\real}{Re}

\begin{document}

\title{Approximative two-flavor framework for neutrino oscillations with
nonstandard interactions}

\author{Mattias Blennow}
\email[]{blennow@mppmu.mpg.de}
\affiliation{Max-Planck-Institut f\"ur Physik,
F\"ohringer Ring 6, 80805 M\"unchen, Germany}
\author{Tommy Ohlsson}
\email[]{tommy@theophys.kth.se}
\affiliation{Department of Theoretical Physics, School of Engineering
Sciences, Royal Institute of Technology (KTH) -- AlbaNova University
Center, Roslagstullsbacken 21, 106 91 Stockholm, Sweden}

\begin{abstract}
In this paper, we develop approximative two-flavor neutrino
oscillation formulas including subleading nonstandard interaction
effects. Especially, the limit when the small mass-squared difference
approaches zero is investigated. The approximate formulas are also
tested against numerical simulations in order to determine their
accuracy and they will probably be most useful in the GeV energy
region, which is the energy region where most upcoming neutrino
oscillation experiments will be operating. Naturally, it is important
to have analytical formulas in order to interpret the physics behind
the degeneracies between standard and nonstandard parameters.
\end{abstract}


\preprint{MPP-2008-45}

\maketitle

\section{Introduction}

Neutrino oscillation physics has had a remarkable progress as the
leading description for neutrino flavor transitions during the last
decade. Even though it serves as the leading description, other
mechanisms could be responsible for transitions on a subleading
level. In this paper, we will develop an approximative framework for
neutrino oscillations with so-called nonstandard interactions (NSIs)
as such subleading effects. This framework is derived in the limit
when the small mass-squared difference is negligible, \ie, $\Delta
m_{21}^2 \to 0$, as well as when the effective matter interaction part
of the Hamiltonian has only one eigenvalue significantly different
from zero. The main effects of NSIs will be parametrized by $v$ and
$\beta$, which will be defined in \Sec~\ref{sec:twoflavor}.

Recently, NSIs with matter have attracted a lot of attention in the
literature. In some sense, the potential for these interactions can be
viewed as a generalization of the coherent forward-scattering
potential, which describes the effects of ordinary (or standard)
interactions with matter \cite{Wolfenstein:1977ue}, and indeed, gives
rise to the famous Mikheyev--Smirnov--Wolfenstein (MSW) effect
\cite{Wolfenstein:1977ue,Mikheev:1986gs,Mikheev:1986wj}. Especially
interesting are approximative frameworks that consider neutrino
oscillations with NSIs which could be used for atmospheric neutrino
data in the $\nu_\mu$-$\nu_\tau$ sector, since these data are
sensitive to interactions of tau neutrinos
\cite{Friedland:2004ah,Friedland:2005vy}. However, there are also
other experimental situations, where one could investigate tau
neutrino interactions such as with the MINOS experiment
\cite{Kitazawa:2006iq,Friedland:2006pi,Blennow:2007pu}. In addition,
in Ref.~\cite{EstebanPretel:2008qi}, the authors have studied an
approximative two-flavor neutrino scenario that is similar to our
framework, but instead used for the OPERA experiment. Nevertheless,
one should be aware of the fact that NSI effects are subleading
effects and have been experimentally constrained
\cite{Davidson:2003ha,Abdallah:2003np}.

The paper is organized as follows. In \Sec~\ref{sec:NSI}, we present
the general scheme for three-flavor neutrino oscillations including
NSIs. Next, in \Sec~\ref{sec:twoflavor}, we develop an approximative
two-flavor neutrino oscillation framework including subleading NSI
effects as well as we compare our framework with earlier
results. Then, in \Sec~\ref{sec:numerics}, we analyze our framework
numerically. In \Sec~\ref{sec:application}, we study applications to
experiments. Finally, in \Sec~\ref{sec:s&c}, we summarize our results
and present our conclusions.

\section{Nonstandard interactions in neutrino oscillations}
\label{sec:NSI}

Standard three-flavor neutrino oscillations in vacuum can be described
by the following Hamiltonian in flavor basis:
\begin{equation}
  H_0 = \frac{1}{2E} U {\rm diag}(0,\Delta m_{21}^2,\Delta m_{31}^2) U^\dagger,
\end{equation}
where $E$ is the neutrino energy, $\Delta m_{ij}^2 \equiv m_i^2 -
m_j^2$ is the mass-squared difference between the $i$th and $j$th mass
eigenstate, and $U$ is the leptonic mixing matrix. By adding the
effective matter Hamiltonian
\begin{equation}
  H_{\rm matter} \simeq {\rm diag}(\sqrt{2} G_F N_e,0,0) = V {\rm
  diag}(1,0,0)
\end{equation}
to the vacuum Hamiltonian $H_0$, standard matter effects on neutrino
oscillations can be studied. In order to investigate the effects of
NSIs between neutrinos and other fermions, another more general
effective interaction Hamiltonian can be added. This effective
Hamiltonian will be of the form
\begin{equation}
  H_{\rm NSI} = V \left( \begin{array}{ccc} \varepsilon_{ee} &
  \varepsilon_{e\mu} & \varepsilon_{e\tau}\\ \varepsilon_{e\mu}^* &
  \varepsilon_{\mu\mu} & \varepsilon_{\mu\tau} \\ \varepsilon_{e\tau}^*
  & \varepsilon_{\mu\tau}^* & \varepsilon_{\tau\tau} \end{array}
  \right),
\end{equation}
where
$$
  \varepsilon_{\alpha\beta} = \sum_{f,a}
	\varepsilon_{\alpha\beta}^{fa} \frac{N_f}{N_e},
$$
$N_f$ is the number of fermions of type $f$, and we have assumed an
unpolarized medium.  Thus, the full Hamiltonian is given by
\begin{equation}
	H = H_0 + H_{\rm matter} + H_{\rm NSI},
\end{equation}
which describes three-flavor neutrino oscillations in matter including NSIs.

\section{Reduction to two-flavor evolution}
\label{sec:twoflavor}

Since the experimental bounds on $\eps_{\mu\alpha}$ and
$\eps_{\alpha\mu}$ are relatively stringent \cite{Davidson:2003ha}, we focus on
the case where the interaction part of the Hamiltonian takes the form
\begin{equation}
	H_{\rm int} = H_{\rm matter} + H_{\rm NSI} = V \mtrx{ccc}{1+\eps_{ee}
	& 0 & \eps_{e\tau} \\ 0 & 0 & 0 \\ \eps_{e\tau}^* & 0 &
	\eps_{\tau\tau}} = V U_{\rm NSI} \diag(v,0,\xi) U_{\rm NSI}^\dagger,
\end{equation}
where the unitary matrix
$$
	U_{\rm NSI} = \mtrx{ccc}{c_\beta & 0 & s_\beta e^{\I\phi} \\ 0 & 1 &
	0 \\ -s_\beta e^{-\I\phi} & 0 & c_\beta}
$$
defines the matter interaction eigenstates, $c_\beta = \cos(\beta)$,
$s_\beta = \sin(\beta)$, and $V$ is the standard MSW potential. In
this framework, the NSIs are parametrized by the two matter
interaction eigenvalues $v$ and $\xi$ as well as the matter
interaction basis parameters $\beta$ and $\phi$ and we have
\begin{equation}
	\eps_{ee} = v c_\beta^2 + \xi s_\beta^2 - 1, \quad
	\eps_{e\tau} = s_\beta c_\beta e^{\I\phi}(\xi-v), \quad
	\eps_{\tau\tau} = v s_\beta^2 + \xi c_\beta^2.
\end{equation}
Thus, the standard neutrino oscillation framework is recovered with
$\beta = \xi = 0$ and $v = 1$.

From the results of atmospheric neutrino experiments, we know that
even high-energy $\nu_\mu$ oscillate, leading to $\xi \ll 1$
\cite{Friedland:2004ah,Friedland:2005vy} for the matter composition of
the Earth. We will here consider the limit $Vv \sim \Delta
m_{31}^2/(2E) \gg V\xi$, which is the applicable limit for standard
neutrino oscillations with energies of a few GeV to tens of GeV inside
the Earth, while earlier papers
\cite{Friedland:2004ah,Friedland:2005vy} have studied the limit $Vv
\gg \Delta m_{31}^2/(2E) \sim V\xi$. In the limit of $\xi \rightarrow
0$ and $\Delta m_{21}^2/(2E) \rightarrow 0$, the full three-flavor
Hamiltonian takes the form (up to an irrelevant addition proportional
to unity)
\begin{equation}
	H = \Delta a_\Delta a_\Delta^\dagger + Vv a_v a_v^\dagger,
\end{equation}
where
$$
	a_\Delta = \mtrx{c}{s_{13}e^{-\I\delta} \\ s_{23}c_{13} \\ c_{23}
	c_{13}} \quad {\rm and} \quad a_v = \mtrx{c}{c_\beta \\ 0 \\ -s_\beta
	e^{-\I\phi}}
$$
are the third column of the leptonic mixing matrix $U$ and first
column of the matter interaction mixing matrix $U_{\rm NSI}$,
respectively. Since this Hamiltonian is defined using only two
linearly independent vectors, $a_\Delta$ and $a_v$, there must be a
third linearly independent vector $a_0$ for which $H a_0 = 0$ and
which is orthogonal to $a_\Delta$ and $a_v$. The components of this
vector are given by
\begin{equation}
	a_{0} = \frac 1{c'}a_\Delta \times a_v,
\end{equation}
where $c' = \cos(\theta')$ is a normalization factor with
$\sin(\theta') = |a_\Delta^\dagger a_v|$ and we have introduced the
product
\begin{equation}
	(A \times B)_i = \eps_{ijk} A^*_j B^*_k,
\end{equation}
which is orthogonal to both $A$ and $B$ as well as antilinear in both
arguments. With our parametrization, we have
\begin{equation}
	s^{\prime 2} = (s_{13}c_\beta - c_{13}s_\beta c_{23})^2 +
	c_{23}\sin(2\beta)\sin(2\theta_{13})\sin^2\left(\frac{\phi+\delta}{2}\right).
	\label{eq:twoflavormix}
\end{equation}
Since $a_0$ is an eigenvector of the full Hamiltonian regardless of
the matter density, its evolution decouples from that of the other two
states and it will only receive a phase factor. Thus, it will be
possible to describe the evolution of the two remaining states in an
effective two-flavor framework. Indeed, if we choose the basis
\begin{equation}
	a_x = e^{\I\alpha_x} a_v, \quad a_y = e^{\I\alpha_y} a_v \times a_0
\end{equation}
(note that $a_y$ is already normalized as $a_v$ and $a_0$ are
orthonormal), with $\alpha_x$ and $\alpha_y$ chosen such that
$a_\Delta^\dagger a_x$ and $a_\Delta^\dagger a_y$ are real, then the
two-flavor Hamiltonian in this basis takes the form
\begin{equation}
	H_2 = V v \mtrx{cc}{1&0\\0&0} + \Delta\mtrx{cc}{s^{\prime 2} & s' c'
	\\ s'c' & c^{\prime 2}},
\end{equation}
which is an ordinary two-flavor Hamiltonian in matter with the vacuum
mixing angle $\theta'$, the mass-squared difference $\Delta m_{31}^2$,
and the MSW potential $Vv$. In order to compute the neutrino flavor
evolution in this basis, any method applicable to two-flavor neutrino
oscillations can be used.

While the above framework makes it easy to reduce the three-flavor
evolution to a two-flavor evolution in the given basis, the final
results have to be projected onto the flavor basis, where
\begin{equation}
	\nu_e = \mtrx{c}{1\\0\\0}, \quad \nu_\mu = \mtrx{c}{0\\1\\0},\quad
	\nu_\tau = \mtrx{c}{0\\0\\1}.
\end{equation}
The two bases are related by a unitary matrix $V$, which is given by
\begin{equation}
	V_{\alpha i} = \nu_\alpha^\dagger a_i.
	\label{eq:2fbasis}
\end{equation}
With this notation and the full evolution matrix given by
\begin{equation}
	S = e^{\I\Phi} \mtrx{ccc}{e^{-\I\Phi} & 0 & 0 \\
	   0 & S_{xx} & S_{xy} \\
	   0 & -S_{xy}^* & S_{xx}^*}
\end{equation}
in the $a_i$ basis, the general neutrino oscillation probability becomes
\begin{eqnarray}
	P_{\alpha\beta} &=&
		|V_{\beta 0}V_{\alpha 0}^*|^2 + (|V_{\beta x}V_{\alpha
	x}^*|^2 + 
		|V_{\beta y}V_{\alpha y}^*|^2)P_{xx} \nonumber \\
	&&
		+ (|V_{\beta x} V_{\alpha y}^*|^2 + |V_{\beta y}
	V_{\alpha x}^*|^2) P_{xy} \nonumber \\
	&&
		+ 2 \real \left[
		V_{\beta0}V_{\alpha 0}^* e^{-\I\Phi}
		\sqrt{P_{xx}}(V_{\beta x}^* V_{\alpha x}
	e^{\I\Lambda_x} + V_{\beta y}^* V_{\alpha y} e^{-\I\Lambda_x}) 
		\right] \nonumber \\
	&& + 2 \real\left[
		V_{\beta0}V_{\alpha 0}^* e^{-\I\Phi}
		\sqrt{P_{xy}}(V_{\beta x}^* V_{\alpha y}
	e^{\I\Lambda_y} - V_{\beta y}^* V_{\alpha x} e^{-\I\Lambda_y}) \right]
		\nonumber \\
	&&
		+2 \real\left[
		P_{xx} V_{\beta x} V_{\alpha x}^* V_{\beta y}^*
	V_{\alpha y} e^{-2\I\Lambda_x} - P_{xy} V_{\beta x} V_{\alpha
	y}^* V_{\beta y}^* V_{\alpha x} e^{-2\I\Lambda_y} \right] \nonumber \\
	&&
		+ 2 \sqrt{P_{xx}P_{xy}} \real\left[
		V_{\beta x} V_{\beta y}^* (
		|V_{\alpha y}|^2 - |V_{\alpha x}|^2
		) e^{-\I(\Lambda_x+\Lambda_y)}
		\right] \nonumber \\
	&&
		+ 2 \sqrt{P_{xx}P_{xy}} \real\left[
		V_{\alpha x}^* V_{\alpha y} (
		|V_{\beta x}|^2 - |V_{\beta y}|^2
		) e^{-\I(\Lambda_x-\Lambda_y)}
		\right],
\end{eqnarray}
where $S_{xx} = \sqrt{P_{xx}}e^{-\I\Lambda_x}$ and $S_{xy} =
\sqrt{P_{xy}} e^{-\I\Lambda_y}$ (with this notation $P_{xx}$ is the
two-flavor survival probability and $P_{xy}$ is the two-flavor
transition probability).

From the above, it becomes apparent that the final neutrino
oscillation probabilities are quite complicated unless further
assumptions are made. This is mainly due to the fact that it is
necessary to project the evolution matrix onto the flavor basis. In
the case of $\beta = 0$, we regain the result from ordinary neutrino
oscillations where $a_v = \nu_e$ and the $\nu_e$ survival probability
takes a very simple form, \ie, $P_{ee} = P_{xx}$. For simplicity, we
present the analytic results only for the case $\delta = \phi = 0$,
where the matrix elements $V_{\alpha i}$ are taken to zeroth order in
$s_{13}$. This simplification will not lead to any large errors, since
$V$ is fixed and $s_{13}$ is small. However, we keep $s_{13} \neq 0$
in the definition of $\theta'$, since there can be resonance effects in
the two-flavor evolution. The resulting neutrino oscillation
probabilities are then given by
\begin{eqnarray}
	c_0^{\prime 4}P_{ee} &=& s_{23}^4 s_\beta^4 + c_\beta^4 \left\{ P_{xx}
	+ 4 c_0^{\prime 2} s_\beta^2 c_{23}^2[P_{xy} \sin^2(\Lambda_y) -
	P_{xx}\sin^2(\Lambda_x)]\right. \nonumber \\ && \left.\phantom{s_{23}^4
	s_\beta^4 +c_\beta^4} + 4\sqrt{P_{xx}P_{xy}} c'_0 s_\beta
	c_{23}(2c_0^{\prime 2}-1)\sin(\Lambda_x) \sin(\Lambda_y)
	\right\} \nonumber
	\\ && + 2s_{23}^2 s_\beta^2 c_\beta^2\left\{ \sqrt{P_{xx}}\left[
	\cos(\Lambda_x - \Phi) - 2s_\beta^2 c_{23}^2 \sin(\Phi) \sin(\Lambda_x)
	\right]\right. \nonumber \\ && \left.\phantom{+ 2s_{23}^2 s_\beta^2
	c_\beta^2} + 2\sqrt{P_{xy}}c'_0 s_\beta c_{23} \sin(\Lambda_y)
	\sin(\Phi)
	\right\}, \label{eq:Pee2f}\\ c_0^{\prime 4}P_{\mu\mu} &=& c_\beta^4
	c_{23}^4 + P_{xx} s_{23}^4 + 2\sqrt{P_{xx}} c_\beta^2 c_{23}^2
	s_{23}^2 \cos(\Lambda_x + \Phi) , \label{eq:Pmm2f} \\ c_0^{\prime 4}
	P_{e\mu} &=& c_\beta^2 s_{23}^2 \left\{ P_{xy} + 2s_\beta^2 c_{23}^2
	[P_{xx} - \sqrt{P_{xx}}\cos(\Lambda_x +\Phi)] \right. \nonumber \\ &&
	\left.\phantom{c_\beta^2 s_{23}^2} +2c'_0 s_\beta
	c_{23}\sqrt{P_{xy}}\left[
	\cos(\Lambda_y+\Phi)-\sqrt{P_{xx}}\cos(\Lambda_x-\Lambda_y) \right]
	\right\}, \label{eq:Pem2f} \\ c_0^{\prime 4} P_{\mu\tau} &=&
	s_{23}^2\left\{ c_\beta^4 c_{23}^2\left[
	\left(1-\sqrt{P_{xx}}\right)^2 +
	4\sqrt{P_{xx}}\sin^2\left(\frac{\Lambda_x+\Phi}{2}\right) \right]
	+s_\beta^2 c_0^{\prime 2} P_{xy} \right.  \nonumber \\ &&
	\left.\phantom{\frac{s_{23}^2}1} +2\sqrt{P_{xy}} c_\beta^2 s_\beta
	c_{23} c' \left[
	\cos(\Lambda_y-\Phi)-\sqrt{P_{xx}}\cos(\Lambda_x+\Lambda_y) \right]
	\right\}, \label{eq:Pmt2f}
\end{eqnarray}
where $c_0^{\prime 2} = 1-s_\beta^2 c_{23}^2$. The five remaining
probabilities can be trivially deduced from the above results using
unitarity conditions. Note that
\eqs~(\ref{eq:Pee2f})--(\ref{eq:Pmt2f}) will be illustrated below in
\figs~\ref{fig:Pmmfig}--\ref{fig:Pmtfig}.

\subsection{Comparison and consistency with earlier results}

As mentioned above, in \Refs~\cite{Friedland:2004ah,Friedland:2005vy},
the limit of $v \rightarrow \infty$ and $\xi V \sim \Delta
m_{31}^2/(2E)$ is studied. They also study the limit of $\xi \rightarrow
0$, which becomes similar to the approach taken in
\Ref~\cite{Blennow:2004js}, where the limit of large matter effects
was studied (the difference being that the former includes NSIs and
the latter includes the effects of nonzero $\theta_{13}$ and $\Delta
m_{21}^2$). In this limit, the eigenstate $a_v$ decouples, and since
$V_{\mu x} = 0$, the $\nu_\mu$ oscillates in a pure vacuum two-flavor
system. It is found that the effective neutrino oscillation parameters
in this system are given by
\begin{eqnarray}
	\Delta m_m^2 &=& \Delta m_{31}^2
	\sqrt{[c_{2\theta}(1+c_\beta^2)-s_\beta^2]^2/4+s_{2\theta}^2c_\beta^2},
	\label{eq:FLeffdm} \\ \tan(2\theta_m) &=& \frac{2s_{2\theta}
	c_\beta}{c_{2\theta}(1+c_\beta^2)-s_\beta^2}, \label{eq:FLefftheta}
\end{eqnarray}
where $c_{2\theta} = \cos(2\theta_{23})$ and $s_{2\theta} =
\sin(2\theta_{23})$.

In our framework, it is easy to obtain the corresponding quantities as
\begin{eqnarray}
	\Delta m_m^2 &=& 2E(a_y^\dagger H a_y - a_0^\dagger H a_0) \nonumber
	 \\ &=& \Delta m_{31}^2 c^{\prime 2} \label{eq:oureffdm},\\
	 \tan(\theta_m) &=& |V_{\mu y}|/|V_{\mu 0}| \nonumber \\ &=&
	 \frac{c_{13}s_{23}}{c_{13}c_\beta c_{23} + s_\beta s_{13}},
	 \label{eq:ourefftheta}
\end{eqnarray}
where the last equality holds only when $\phi = \delta = 0$ and
becomes more complicated when this is not the case. It is easy to show
that these results are the same as those of
\Refs~\cite{Friedland:2004ah,Friedland:2005vy} in the limit of
$\theta_{13} \rightarrow 0$, where we obtain
\begin{eqnarray}
	\Delta m_m^2 &=& \Delta m_{31}^2 (1-s_\beta^2 c_{23}^2),
	\label{eq:dm2eff0} \\
	\tan(\theta_m) &=& \frac{\tan(\theta_{23})}{c_\beta},
	\label{eq:thetaeff0}
\end{eqnarray}
which are slightly simpler forms of \eqs~(\ref{eq:FLeffdm}) and
(\ref{eq:FLefftheta}).  Note that \eqs~(\ref{eq:oureffdm}) and
(\ref{eq:ourefftheta}) are valid for any value of $\theta_{13}$,
whereas \eqs~(\ref{eq:FLeffdm}) and (\ref{eq:FLefftheta}) are only
valid in the limit $\theta_{13} \rightarrow 0$.  Our results also
agree with those of \Ref~\cite{Blennow:2004js} in the limit of no NSIs
and $\Delta m_{21}^2 \ll \Delta m_{31}^2$.

\section{Numeric analysis of oscillation probabilities}
\label{sec:numerics}

In this section, we present numerical results to show the accuracy of
the two-flavor approximation presented in the previous section. In
order to perform this, we will present the neutrino oscillation
probabilities in terms of neutrino oscillograms of the Earth, \ie,
probability iso-contours in the
zenith-angle--neutrino-energy plane. Since we are mainly interested
in the accuracy of the approximation (rather than making quantitative
predictions), we use a simplified Earth model, where the mantle and
core are approximated as having constant matter densities (4.65
g/cm$^3$ and 10.2 g/cm$^3$, respectively). For the standard neutrino
oscillation parameters, we have used the values given in
\tab~\ref{tab:oscparams}.
\begin{table}
\begin{tabular}{lc}
\hline\hline
Parameter & Value \\
\hline
$\Delta m_{31}^2$ & $2.7\cdot 10^{-3}$~eV$^2$ \\
$\Delta m_{21}^2$ & $8\cdot 10^{-5}$~eV$^2$ \\
$\theta_{12}$ & $33.2^\circ$ \\
$\theta_{23}$ & $45^\circ$ \\
$\theta_{13}$ & 0 or 8$^\circ$ \\
\hline \hline
\end{tabular}
\caption{The neutrino oscillation parameters used in our numerical
examples. The value of $\theta_{13}$ is specified in each figure. In
addition, for all examples, we have used $\delta = 0$.}
\label{tab:oscparams}
\end{table}

As will be observed in \figs~\ref{fig:Pmmfig}--\ref{fig:Pmtfig}, the
approximations of \eqs~(\ref{eq:Pee2f})--(\ref{eq:Pmt2f}) are
surprisingly accurate in describing the shapes of the oscillograms,
the main source of error coming from the negligence of the small
mass-squared difference $\Delta m_{21}^2$ (putting $\Delta m_{21}^2 =
0$ in the numerical simulations would lead to differences between the
approximations and the numerical results which would be hardly
noticeable).

We will also observe that, in all cases, the NSI terms dominate the
high-energy behavior of the neutrino oscillation probabilities, as
expected. However, in addition, it will be apparent that the effective
two-flavor mixing angle $\theta'$ plays a large role in determining
the qualitative behavior at energies around the resonance energy. This
includes discussions on where the resonance appears (both in energy
and baseline), as well as on other resonant effects such as parametric
resonance
\cite{Krastev:1989ix,Liu:1998nb,Petcov:1998su,Akhmedov:1998ui,Akhmedov:1998xq,Akhmedov:2005yj}.
For energies lower than the ones displayed in
\figs~\ref{fig:Pmmfig}--\ref{fig:Pmtfig}, our approximation will
become worse due to the fact that the effects of $\Delta m_{21}^2$
will no longer be negligible.

\subsection{The muon neutrino survival probability $\boldsymbol{P_{\mu\mu}}$}

\begin{figure}
	\begin{center}
		\includegraphics[width=0.9\textwidth,clip=true]{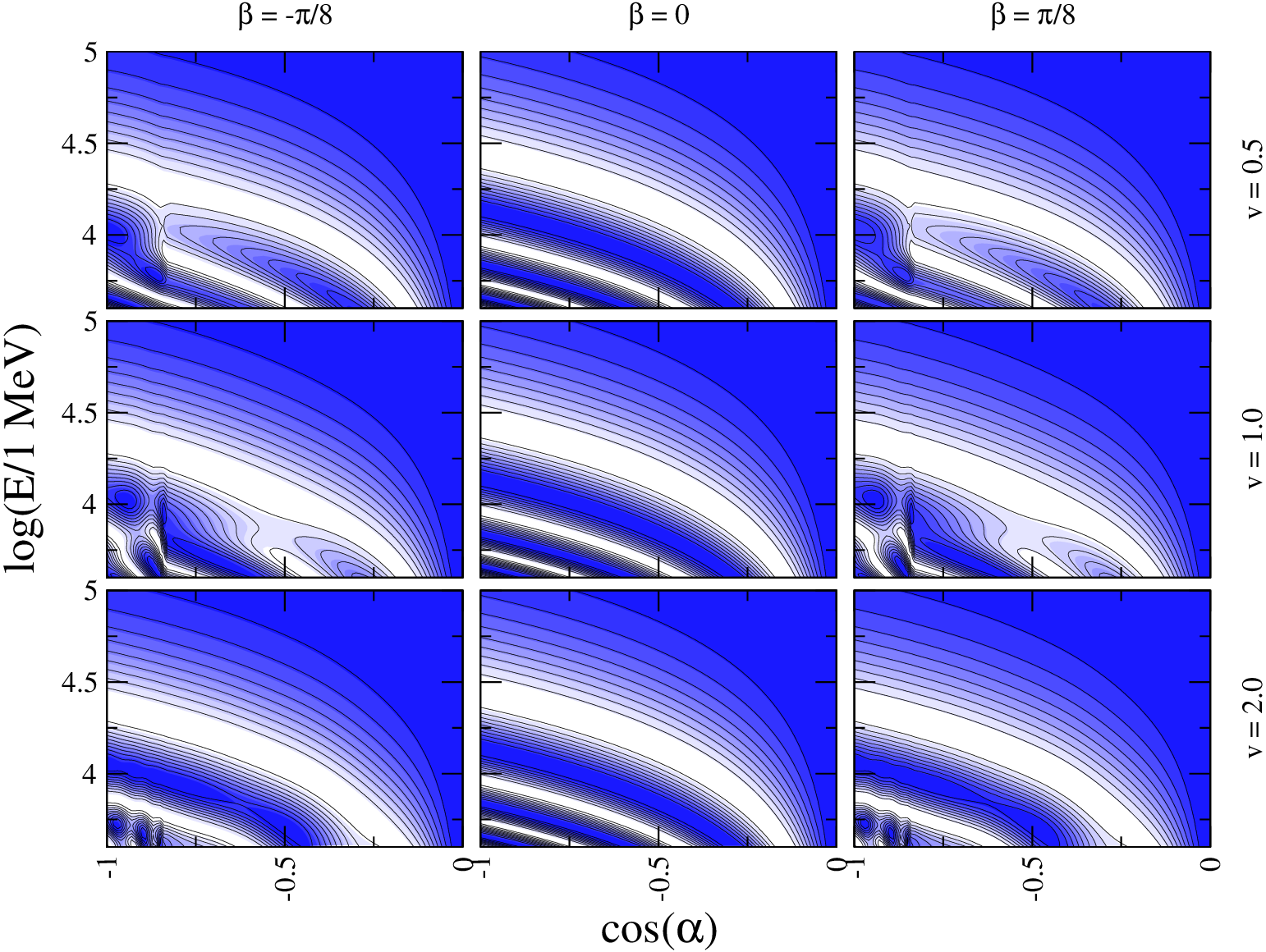}
		\caption{The muon neutrino survival probability
		$P_{\mu\mu}$ as a function of zenith angle and energy
		for different values of $v$ and $\beta$. The colored
		regions correspond to the numerical results (darker
		regions for higher probability), while the solid black
		curves correspond to the two-flavor approximation
		introduced in \Sec~\ref{sec:twoflavor}. In this
		figure, the leptonic mixing angle $\theta_{13}$ is set
		to zero.}
		\label{fig:Pmmfig}
	\end{center}
\end{figure}

In the next section, we will study the muon neutrino disappearance
channel at a neutrino factory using the General Long Baseline
Experiment Simulator (GLoBES). Thus, our first numerical example,
shown in \fig~\ref{fig:Pmmfig}, is the muon neutrino survival
probability $P_{\mu\mu}$.
As was discussed earlier, at high neutrino energies, the muon neutrino
will oscillate in a pure two-flavor vacuum scenario with the effective
parameters given by \eqs~(\ref{eq:dm2eff0}) and (\ref{eq:thetaeff0}),
since we have used $\theta_{13} = 0$ in the construction of
\fig~\ref{fig:Pmmfig}. Furthermore, the choice of $\theta_{13} = 0$
and the approximation that $\Delta m_{21}^2$ is small imply that the
low-energy neutrinos also oscillate in a two-flavor vacuum scenario,
but with the parameters $\Delta m^2 = \Delta m_{31}^2$ and $\theta =
\theta_{23}$. In the case of $\beta = 0$, these two cases are
obviously equivalent and the matter interaction effects do not change
the situation at any energy (since the $\nu_e$-$\nu_e$ element is the
only nonzero element of the interaction contribution to the
Hamiltonian). However, for $\beta \neq 0$, there will be resonance
effects in the region where $\Delta \sim Vv$. As can be seen from
\fig~\ref{fig:Pmmfig}, these resonance effects are present mainly
around $E \simeq 10^{3.7}$~MeV for $v = 1$ and at energies
appropriately displaced from this for $v = 0.5$ and $v = 2$ (\ie, $E
\simeq 10^{4}$~MeV and $E \simeq 10^{3.4}$~MeV, respectively). Note
that the colored regions in \figs~\ref{fig:Pmmfig}--\ref{fig:Pmtfig}
represent the ``exact'' (or numerical) three-flavor results, whereas
the black solid curves are the approximate two-flavor results given by
\eqs~(\ref{eq:Pee2f})--(\ref{eq:Pmt2f}). In general, these two-flavor
results are in excellent agreement with the exact three-flavor
results, and therefore, they can be used as good approximations in
analyses of neutrino data.

\subsection{The electron neutrino survival probability $\boldsymbol{P_{ee}}$}

\begin{figure}
	\begin{center}
		\includegraphics[width=0.75\textwidth,clip=true]{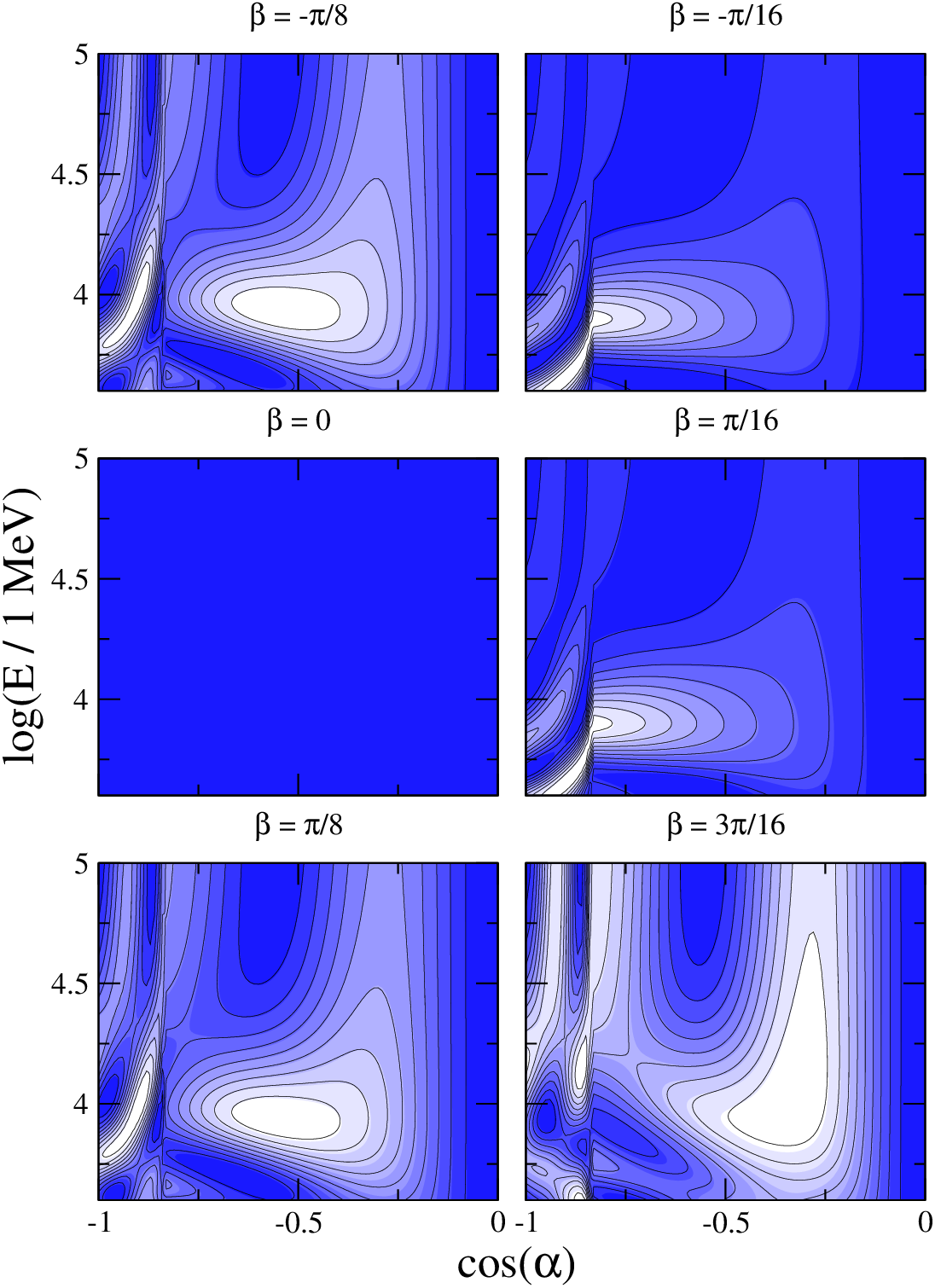}
		\caption{The electron neutrino survival probability
		$P_{ee}$ as a function of zenith angle and energy for
		$\theta_{13} = 0$, $v = 1$, and different values of
		$\beta$. The colored regions and solid black curves
		are constructed as in \fig~\ref{fig:Pmmfig}.}
		\label{fig:Pee0fig}
	\end{center}
\end{figure}

\begin{figure}
	\begin{center}
		\includegraphics[width=0.75\textwidth,clip=true]{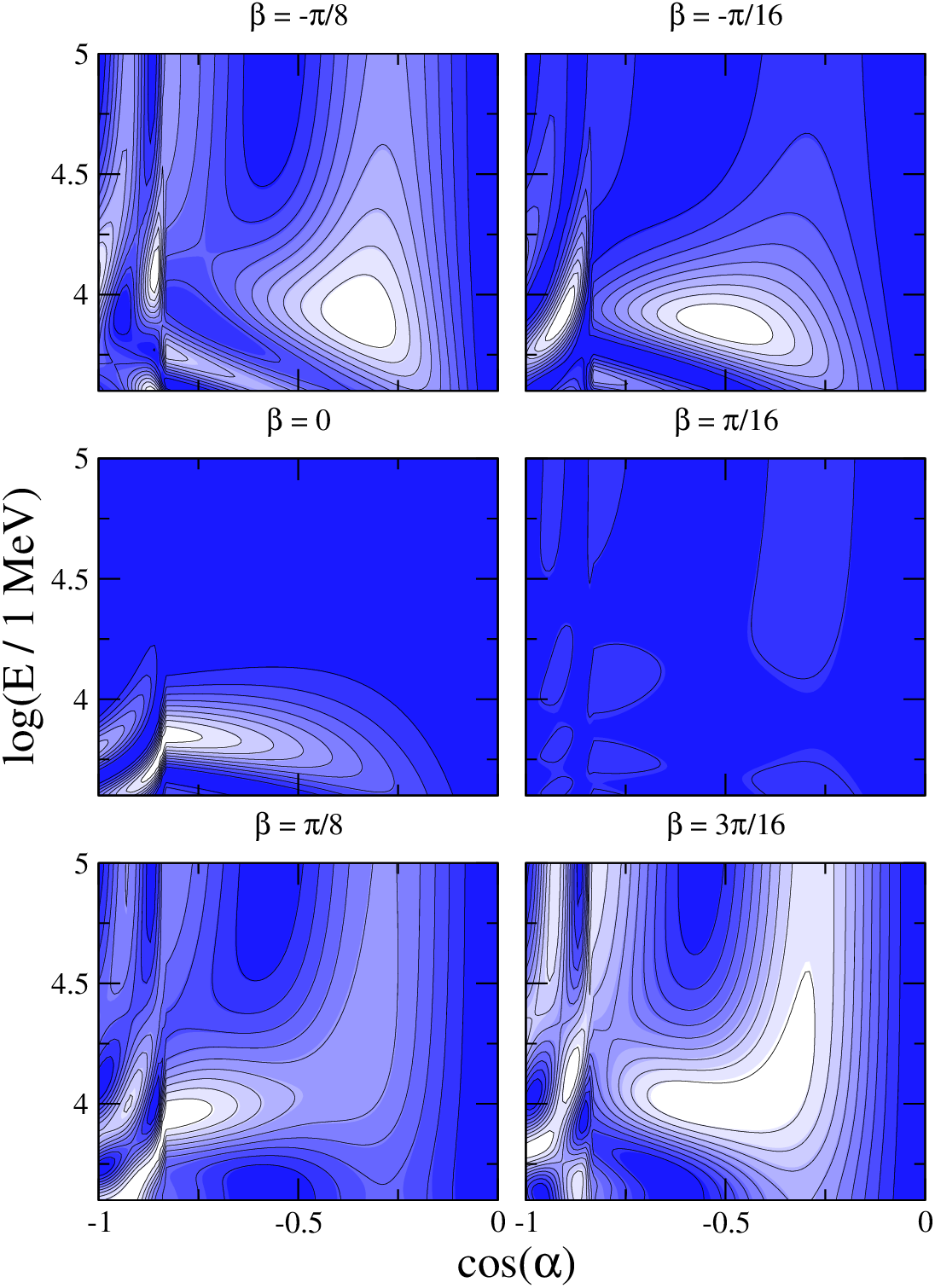}
		\caption{The same as \fig~\ref{fig:Pee0fig}, but for
		$\theta_{13} = 8^\circ$.}
		\label{fig:Pee8fig}
	\end{center}
\end{figure}

As discussed in \Ref~\cite{Akhmedov:2006hb}, the electron neutrino
survival probability $P_{ee}$ can be a very useful tool when
considering ways of determining
the small leptonic mixing matrix element $U_{e3}$. However, as has
been discussed earlier in this paper (and also in previous papers
\cite{Huber:2001de,Blennow:2005qj,Friedland:2006pi,Kitazawa:2006iq,Blennow:2007pu}), the presence of NSIs can significantly alter the
interpretation of a positive oscillation signal in this channel. In
particular, the effective two-flavor mixing angle $\theta'$ is given
by \eq~(\ref{eq:twoflavormix}) and the effective matrix element
$\tilde U_{e3}$ by \cite{Blennow:2007pu}
\begin{equation}
	\tilde U_{e3} \simeq U_{e3} + \eps_{e\tau} \frac{2EV}{\Delta
	m_{31}^2}c_{23}.
\end{equation}
In \figs~\ref{fig:Pee0fig} and \ref{fig:Pee8fig}, we present the
oscillograms for $P_{ee}$ using different values of $\beta$ and with
$\theta_{13} = 0$ and $8^\circ$, respectively.
It should be noted that $s' \simeq 0$ in the case when $\theta_{13} =
8^\circ$ and $\beta = \pi/16$ (this is of course the reason for
choosing $\theta_{13} = 8^\circ$). As can be observed in these two
figures, the high-energy ($E \gtrsim 10^{4.5}$~MeV) behavior of the
survival probability $P_{ee}$ is practically unaffected by the change
in $\theta_{13}$. This is clearly to be expected, since the vacuum
neutrino oscillation terms are suppressed by the neutrino energy $E$,
while the matter interaction terms remain constant. Simply neglecting
the vacuum oscillation terms, the expectation at high energy is
\begin{equation}
	1 - P_{ee} = \sin^2(2\beta) \sin^2(VvL),
\end{equation}
which can also be obtained as a limiting case of \eq~(\ref{eq:Pee2f}).

However, for energies in the resonance regime, the qualitative
behavior of $P_{ee}$ is clearly dominated by the value of the
effective two-flavor mixing angle $\theta'$. This is particularly
apparent in the mid-row panels of \figs~\ref{fig:Pee0fig} and
\ref{fig:Pee8fig}. While the oscillation probability in the resonance
region is kept small in both the mid-left panel of
\fig~\ref{fig:Pee0fig} and the mid-right panel of
\fig~\ref{fig:Pee8fig} (corresponding to $\theta' \simeq 0$), the
resonant behavior at $E \sim 10^{3.8}$~MeV in the mid-right panel of
\fig~\ref{fig:Pee0fig} and the mid-left panel of
\fig~\ref{fig:Pee8fig} (both corresponding to $\theta' \simeq
8^\circ$) are also very similar.

\subsection{The neutrino oscillation probability $\boldsymbol{P_{e\mu}}$}

\begin{figure}
	\begin{center}
		\includegraphics[width=0.75\textwidth,clip=true]{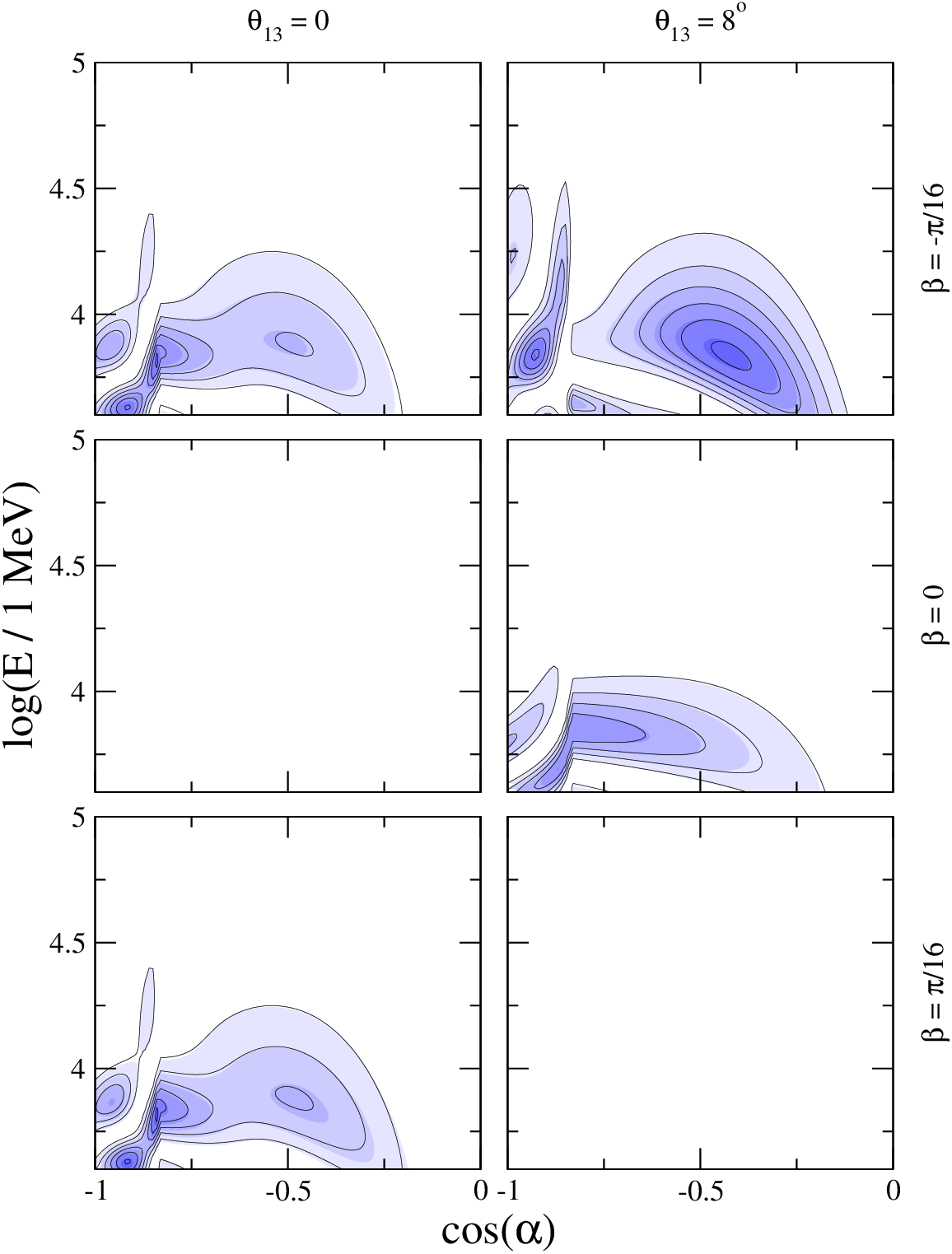}
		\caption{The neutrino oscillation probability
		$P_{e\mu}$ as a function of zenith angle and energy
		for different values of $\theta_{13}$ and $\beta$. For
		this figure, we have used $v = 1$ and the colored
		regions and solid black curves are constructed as in
		\fig~\ref{fig:Pmmfig}.}
		\label{fig:Pemfig}
	\end{center}
\end{figure}

The oscillation of $\nu_e$ into $\nu_\mu$ at neutrino factories has
been thoroughly discussed in several papers (see
\Ref~\cite{Bandyopadhyay:2007kx} and references therein) and is
commonly known as the ``golden'' channel. Thus, in
\fig~\ref{fig:Pemfig}, we present oscillograms for the neutrino
oscillation probability $P_{e\mu}$ for different values of
$\theta_{13}$ and $\beta$.
Comparing this figure with \fig~\ref{fig:Pee8fig}, we can observe
that, for $\theta_{13} = 8^\circ$ and $\beta = 0$, half of the
oscillating $\nu_e$ oscillates into $\nu_\mu$. This is a direct result
of the leptonic mixing angle $\theta_{23}$ being maximal and has been
known for a long time (see, \eg,
\Ref~\cite{Akhmedov:1999uz}). However, this is no longer true when
$\beta \neq 0$. For the two top panels and the lower-left panel, there
are regions where $P_{e\mu} > 0.5$; thus indicating that more than
half of the $\nu_e$ oscillates into $\nu_\mu$ in these regions. It
should be noted that there is an important caveat, if $\theta_{23}$
deviates from maximal mixing, the linear combination of $\nu_\mu$ and
$\nu_\tau$, which $\nu_e$ oscillates into in the standard framework
with nonzero $\theta_{13}$, is not an equal mixture. Therefore, even
the standard neutrino oscillation framework can allow for $P_{e\mu} >
0.5$. However, what is important to note is that the ratio $r =
P_{e\mu}/P_{e\tau}$ is constant in the standard framework, but depends
on both baseline and energy when $\beta \neq 0$. For example, if we
consider neutrino oscillations with $\theta_{13} = 8^\circ$ and $\beta
= -\pi/16$ at $\cos(\alpha) \simeq 0.3$, then $r = 0$ at high energies
due to the decoupling of the $\nu_\mu$ when the matter interaction
terms dominate, while $r > 1$ in the relatively low-energy region
around the resonance. Hence, measuring the energy and baseline
dependence of $r$ could be a way of determining whether NSIs are
present as long as the small mass-squared difference $\Delta m_{21}^2$
is negligible. As in the case of the neutrino survival probability
$P_{ee}$, we can observe that there are practically no oscillations in
the resonance region for the cases where $\theta' \simeq 0$.

\subsection{The neutrino oscillation probability $\boldsymbol{P_{\mu\tau}}$}

\begin{figure}
	\begin{center}
		\includegraphics[width=0.75\textwidth,clip=true]{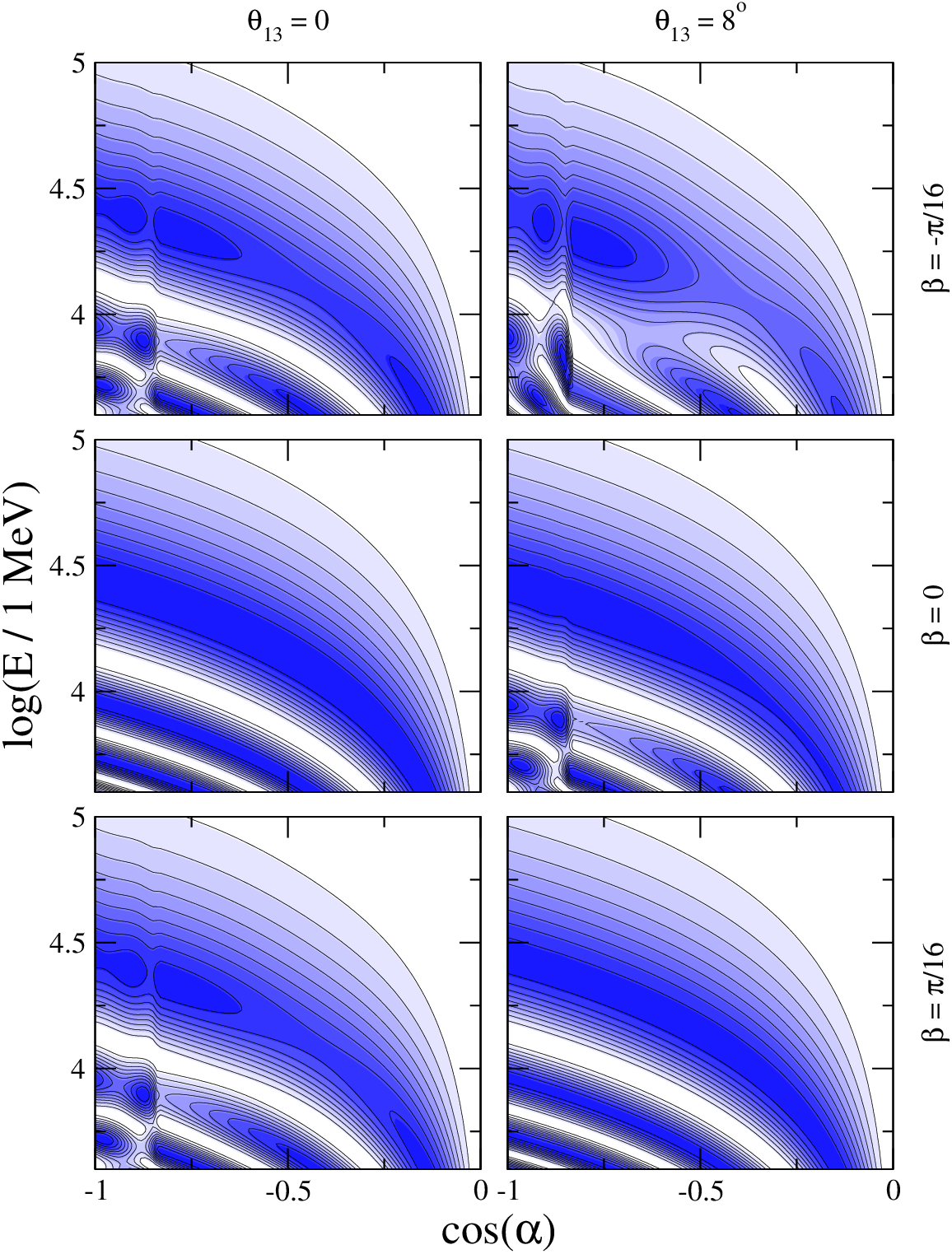}
		\caption{The neutrino oscillation probability
		$P_{\mu\tau}$ as a function of zenith angle and energy
		for different values of $\theta_{13}$ and $\beta$. For
		this figure, we have used $v = 1$ and the colored
		regions and solid black curves are constructed as in
		\fig~\ref{fig:Pmmfig}.}
		\label{fig:Pmtfig}
	\end{center}
\end{figure}

In order to accommodate all of \eqs~(\ref{eq:Pee2f})--(\ref{eq:Pmt2f})
in our numerical study, we also show the results for the neutrino
oscillation probability $P_{\mu\tau}$ in \fig~\ref{fig:Pmtfig}.
Again, we can observe the fact that the oscillations for given values
of $\theta'$ turn out to be very similar. Especially, the panels
where $\theta' \simeq 0$ are almost identical and very similar to pure
vacuum neutrino oscillations. In particular, the effects of going from
mantle-only trajectories to core-crossing trajectories at about
$\cos(\alpha) \simeq 0.84$ is not apparent in these figures, the
reason being the fact that the Hamiltonian is diagonalized by the
matrix $(V_{\alpha i})$ given in \eq~(\ref{eq:2fbasis}) for all values
of $VE$.

It should also be noted that $P_{\mu e} = P_{e\mu}$, since we use a
symmetric matter density profile and no \CP-violating phases in this
section. Thus, the differences between the figures for
the $P_{\mu\mu}$ probabilities and those for the $P_{\mu\tau}$
probabilities (except for the interchange of dark and light areas) are
summarized in the figures for the $P_{e\mu}$ probabilities as $P_{\mu
e} + P_{\mu\mu} + P_{\mu\tau} = 1$.

\section{Application to a fictive neutrino factory}
\label{sec:application}

In this section, we study the impact of NSIs on the sensitivity
contours of a fictive neutrino factory experiment. In order to achieve
this, we have used the General Long-Baseline Experiment Simulator
(GLoBES) \cite{Huber:2004ka,Huber:2007ji} and a modified version of
the standard neutrino factory experimental setup included in the
GLoBES distribution (essentially ``NuFact-II'' from
\Ref~\cite{Huber:2002mx}), modified to only take the $\nu_\mu$
disappearance channel into account and to have a baseline of $L =
7000$~km (from \fig~\ref{fig:Pmmfig}, we note that the NSI effects
should be relatively strong at this baseline, it is also very close to
the ``magic'' baseline \cite{Huber:2003ak}, which means that the
impact of the solar parameters is expected to be small). This neutrino
factory setup corresponds to a parent muon energy of 50~GeV and an
assumed target power of 4~MW. The running time is set to four years in
each polarity and the detector is assumed to be a magnetized iron
calorimeter with a fiducial mass of 50~kton. For the neutrino
cross-sections, we have also used the files included in the GLoBES
distribution, which are based on
\Refs~\cite{Messier:1999kj,Paschos:2001np}. We consider neutrino
energies in the range 4--50~GeV, divided into 20 equally spaced energy
bins.  The simulated neutrino oscillation parameters are the same as
in the previous section, \ie, they are given in
\tab~\ref{tab:oscparams}. The simulated $\theta_{13}$ has been put to
zero for simplicity.

\subsection{Determination of standard parameters}

One important question when considering NSIs is how the inclusion of
NSIs can alter the sensitivity of an experiment. This subject has been
studied for a number of different neutrino oscillation experiments
\cite{Huber:2001de,Friedland:2004ah,Friedland:2005vy,Blennow:2005qj,Friedland:2006pi,Blennow:2007pu,EstebanPretel:2008qi,Kopp:2008ds,Blennow:2008ym}. At
the neutrino factory in question, the $\nu_\mu$ disappearance channel
will normally be a very sensitive probe to the neutrino oscillation
parameters $\Delta m_{31}^2$ and $\theta_{23}$. However, with the
inclusion of NSIs, there is an additional degeneracy, leading to
experiments becoming less sensitive to the standard neutrino
oscillation parameters. Note that this degeneracy has been
investigated before in the case of atmospheric neutrinos in
\Refs~\cite{Friedland:2004ah,Friedland:2005vy}. In particular, the
major part of the neutrino energies at the neutrino factory described
before are well above the resonance energy. Thus, the induced
degeneracy in the $\sin^2(\theta_{23})$--$\Delta m_{31}^2$ plane
should be well described by \eqs~(\ref{eq:oureffdm}) and
(\ref{eq:ourefftheta}).

\begin{figure}
	\begin{center}
		\includegraphics[width=0.8\textwidth,clip=true]{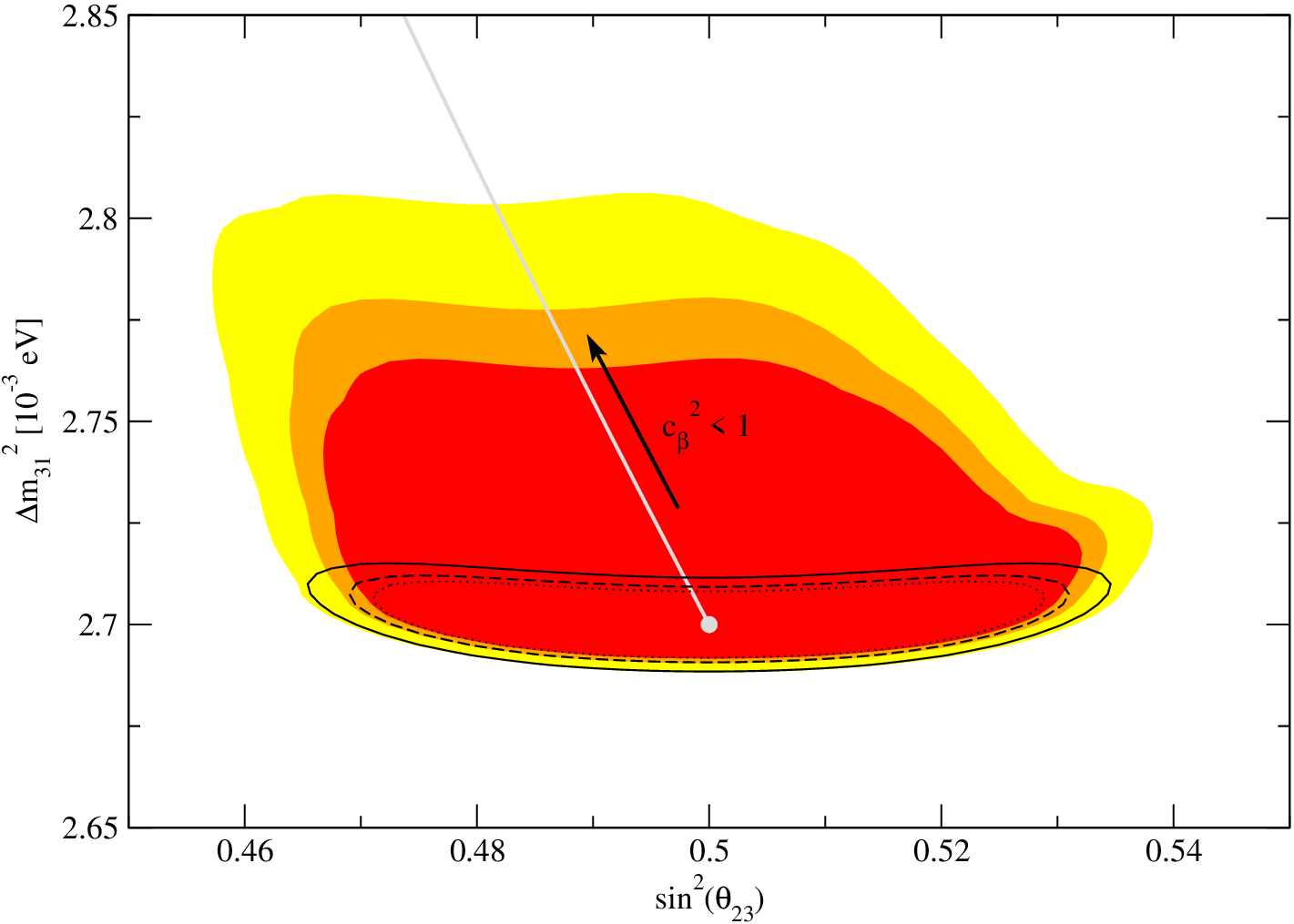}
		\caption{The sensitivity contours (2~d.o.f.) in the
		$\sin^2(2\theta_{23})$--$\Delta m_{31}^2$ plane for
		the $\nu_\mu$ disappearance channel at the fictive
		neutrino factory experiment described in the text. The
		black curves correspond to the sensitivity if it is
		assumed that there are no NSIs present, while the
		colored regions correspond to the sensitivity if NSIs
		within current bounds are taken into account. The gray
		circle corresponds to the simulated parameters and the
		gray curve to the expected direction of the NSI
		degeneracy. The confidence levels are 90~\%, 95~\%,
		and 99~\%, respectively. We have assumed Gaussian
		priors of $\eps_{ee} < 2.6$, $\eps_{e\tau} < 1.9$, and
		$\eps_{\tau\tau} < 1.9$ at a 90~\% confidence level.}
		\label{fig:s23dm2fig}
	\end{center}
\end{figure}

In \fig~\ref{fig:s23dm2fig}, we can observe how this degeneracy
manifests itself in the sensitivity of the $\nu_\mu$ disappearance
channel.
While the sensitivity contours are symmetric around
$\sin^2(\theta_{23}) = 0.5$ when not including NSIs (\ie, $\beta =
0$), they extend to smaller values of $\theta_{23}$ and larger values
of $\Delta m_{31}^2$ when NSIs are included (\ie, $\beta \neq 0$). The
conclusion that the sensitivity contours will extend to smaller
$\theta_{23}$ and larger $\Delta m_{31}^2$ is general and apparent
from the form of \eqs~(\ref{eq:oureffdm}) and (\ref{eq:ourefftheta}),
where the same $\theta_m$ and $\Delta m_m^2$ can be achieved for a
class of different $\theta_{23}$ and $\Delta m_{31}^2$ given by
\begin{eqnarray}
\tan(\theta_{23}) &=& c_\beta \tan(\theta_m), \\
\Delta m_{31}^2 &=& \frac{\Delta m_m^2}{1-s_\beta^2 c_{23}^2},
\end{eqnarray}
in the case when $\theta_{13} = 0$. This degeneracy has also been
marked in \fig~\ref{fig:s23dm2fig} for $\theta_{23}$ and $\Delta
m_{31}^2$ equal to the simulated values when $\beta = 0$.

\subsection{Determination of NSI parameters}

\begin{figure}
	\begin{center}
		\includegraphics[width=0.8\textwidth,clip=true]{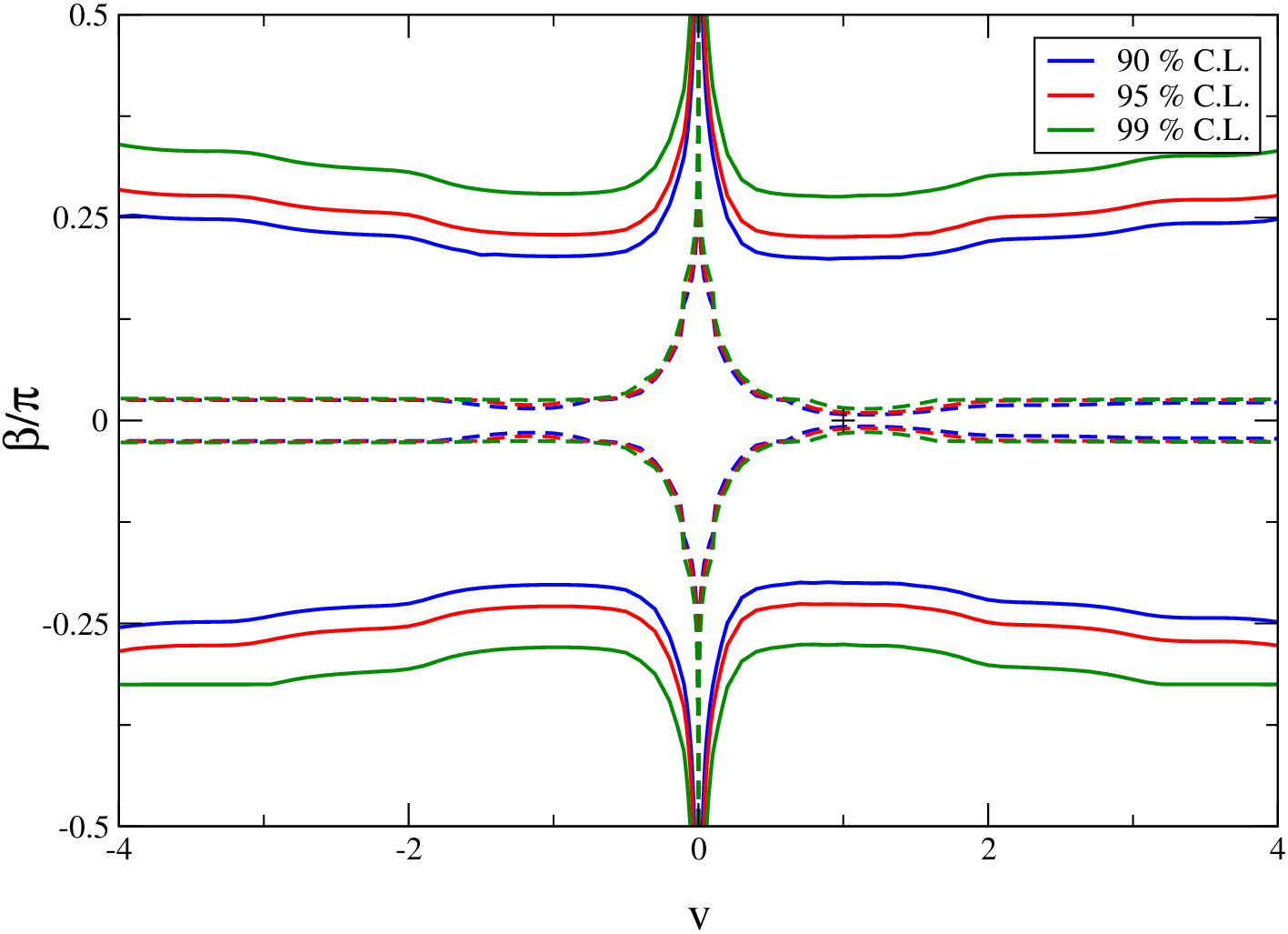}
		\caption{The sensitivity (2~d.o.f.) to the NSI
		parameters $v$ and $\beta$ for the $\nu_\mu$
		disappearance channel at the fictive neutrino factory
		experiment described in the text. The dashed curves
		are the systematics-only sensitivities and the solid
		curves include parameter correlations with the
		standard neutrino oscillation parameters (with errors
		comparable to the current errors). The contours were
		constructed assuming no NSIs and the $+$ corresponds
		to this situation.}
		\label{fig:NSIfig}
	\end{center}
\end{figure}

Another interesting question in relation to NSIs and future
experiments is what the experimental sensitivity to NSIs is, \eg, what
bounds the experiments could put on the NSI parameters. Thus, in
\fig~\ref{fig:NSIfig}, we show the sensitivity contours in the
$v$--$\beta$ plane (the other NSI parameters have been marginalized)
corresponding to the $\nu_\mu$ disappearance channel at a neutrino
factory described above.
From this figure, we notice that the given experiment is not very
sensitive to the NSI parameters as long as the knowledge on the
standard neutrino oscillation parameters has errors which are
comparable to those of the current experimental bounds. Essentially,
for large $v$ (such that the entire energy spectrum is well above
resonance), the bound on $\beta$ comes from the mismatch between the
effective $\theta_m$ and $\Delta m_m^2$ measured by the neutrino
factory and the external measurements on $\theta_{23}$ and $\Delta
m_{31}^2$ and the errors of both are propagated to the error in the
determination of $\beta$. However, if the external errors on the
standard neutrino oscillation probabilities are very small (\ie, such
that we do not need to consider correlations with the standard
parameters), then the bounds on the NSI parameters are significantly
improved. It should be noted that the insensitivity to $\beta$ for
small $|v|$ is clearly due to the fact that it is not possible to
determine $\beta$ when the interaction effects are suppressed. We
should also note that the given experiment is clearly not the best to
use in order to distinguish NSIs from standard neutrino
oscillations. For example, it should be more viable to detect NSIs
through other oscillation channels or through the combination of the
$\nu_\mu$ disappearance channel with one of the appearance channels.

\section{Summary and conclusions}
\label{sec:s&c}

We have considered the effects of NSIs on the propagation of
$\geq$~GeV neutrinos in the Earth with the assumption that $\Delta
m_{21}^2$ is small and that $\eps_{ee}$, $\eps_{e\tau}$, and
$\eps_{\tau\tau}$ are the only important NSI parameters. We have shown
that, if the NSI parameters are constrained to be on the ``atmospheric
parabola,'' then the neutrino flavor evolution matrix can be
straightforwardly computed using a simple two-flavor scenario in
matter with a mixing angle given by \eq~(\ref{eq:twoflavormix}). This
two-flavor scenario is the natural NSI extension of the usual
two-flavor approximation used in standard neutrino oscillations when
$\Delta m_{21}^2$ is small. While this two-flavor scenario, unlike the
standard two-flavor scenario, is not diagonal in flavor space, it
should be noted that if NSI effects are present in the propagation,
then they should also be present in the creation and detection
processes, meaning that the actual flavor space is no longer
equivalent to the weak interaction flavor space. Deviations
from the atmospheric parabola or other NSI parameters can be easily
treated as a perturbation to the two-flavor scenario.

The impact of NSIs on neutrino oscillations in the Earth as well as
the accuracy of our two-flavor approximation have been studied using
neutrino oscillograms of the Earth. It was noted that the resonance
regions in the neutrino survival channels are mainly dependent on the
mixing angle of the effective two-flavor scenario, implying a
degeneracy between $\theta_{13}$ and the NSI angle $\beta$. This
degeneracy is preferentially broken by using data from above or below
the resonance, or by studying the flavor combination into which
neutrinos actually oscillate.

Finally, we have given an example of the impact of the NSIs in a
fictive neutrino factory with a baseline of $L = 7000$~km studying the
muon neutrino disappearance channel. We have shown that the degeneracy
between the standard and NSI parameters extends in the expected
direction and computed the actual sensitivity to the NSI parameters,
both by assuming no degeneracy with the standard parameters (\ie, the
standard parameters assumed to be exactly known) and by assuming
correlations with the standard parameters (in this case, Gaussian
priors with a size corresponding to the current experimental errors
were used). Again, it should be stressed that this part is only
intended to show the impact of the NSI parameters in a fictive
experiment and does not offer a realistic approach for searching for
NSIs. Such considerations can be found in \Ref~\cite{Kopp:2008ds}.

\begin{acknowledgments}

We would like to thank Evgeny Akhmedov for useful discussions. This
work was supported by the Swedish Research Council
(Vetenskapsr\aa{}det), Contract Nos.~623-2007-8066 (M.B.) and
621-2005-3588 (T.O.) and the Royal Swedish Academy of Sciences (KVA)
(T.O.).

\end{acknowledgments}

\end{document}